\documentclass[%
 reprint,
superscriptaddress,
twocolumn,
 amsmath,amssymb,
 aps,
pra,
longbibliography,
]{revtex4-2}

\usepackage{graphicx}
\usepackage{dcolumn}
\usepackage{bm}
\usepackage{xcolor}
\usepackage{float}
\usepackage{soul}
\usepackage{natbib}

\usepackage{amscd} 
\usepackage{amsfonts} 
\usepackage{amssymb} 
\usepackage{latexsym} 

\begin{document}

\preprint{APS/123-QED}

\title{All-Optical Experimental Control of High-Harmonic Photon Energy}

\author{L\'{e}n\'{a}rd Guly\'{a}s Oldal}
\thanks{These authors contributed equally to this work}
\affiliation{ELI-ALPS, ELI-HU Non-Profit Ltd., Wolfgang Sandner utca 3., H-6728 Szeged, Hungary}
\affiliation{email: lenard.gulyas@eli-alps.hu}

\author{Peng Ye}
\thanks{These authors contributed equally to this work}
\affiliation{ELI-ALPS, ELI-HU Non-Profit Ltd., Wolfgang Sandner utca 3., H-6728 Szeged, Hungary}

\author{Zolt\'{a}n Filus}
\affiliation{ELI-ALPS, ELI-HU Non-Profit Ltd., Wolfgang Sandner utca 3., H-6728 Szeged, Hungary}

\author{Tam\'{a}s Csizmadia}
\affiliation{ELI-ALPS, ELI-HU Non-Profit Ltd., Wolfgang Sandner utca 3., H-6728 Szeged, Hungary}

\author{T\'{i}mea Gr\'{o}sz}
\affiliation{ELI-ALPS, ELI-HU Non-Profit Ltd., Wolfgang Sandner utca 3., H-6728 Szeged, Hungary}

\author{Massimo De Marco}
\affiliation{ELI-ALPS, ELI-HU Non-Profit Ltd., Wolfgang Sandner utca 3., H-6728 Szeged, Hungary}

\author{Zsolt Bengery}
\affiliation{ELI-ALPS, ELI-HU Non-Profit Ltd., Wolfgang Sandner utca 3., H-6728 Szeged, Hungary}

\author{Imre Seres}
\affiliation{ELI-ALPS, ELI-HU Non-Profit Ltd., Wolfgang Sandner utca 3., H-6728 Szeged, Hungary}

\author{Barnab\'{a}s Gilicze}
\affiliation{ELI-ALPS, ELI-HU Non-Profit Ltd., Wolfgang Sandner utca 3., H-6728 Szeged, Hungary}

\author{P\'{e}ter J\'{o}j\'{a}rt}
\affiliation{ELI-ALPS, ELI-HU Non-Profit Ltd., Wolfgang Sandner utca 3., H-6728 Szeged, Hungary}

\author{Katalin Varj\'{u}}
\affiliation{ELI-ALPS, ELI-HU Non-Profit Ltd., Wolfgang Sandner utca 3., H-6728 Szeged, Hungary}

\author{Subhendu Kahaly}
\affiliation{ELI-ALPS, ELI-HU Non-Profit Ltd., Wolfgang Sandner utca 3., H-6728 Szeged, Hungary}

\author{Bal\'{a}zs Major}
\affiliation{ELI-ALPS, ELI-HU Non-Profit Ltd., Wolfgang Sandner utca 3., H-6728 Szeged, Hungary}

\date{\today}

\begin{abstract}
We generate high-order harmonics in gaseous medium with tunable photon energy using time domain interferometry of double pulses in a non-collinear generation geometry. The method is based on the fact that the generated harmonics inherit certain spectral properties of the driving laser. The two temporally delayed ultrashort laser pulses, identical in all parameters, are produced by a custom-made split-and-delay unit utilizing wave front splitting without a significant energy loss. The arrangement is easy to implement in any attosecond pulse generation beamline, and is suitable for the production of an extreme ultraviolet source with simply and quickly variable central photon energy, useful for a broad range of applications.
\end{abstract}

\maketitle

Because of the continuous development of ultrafast laser technology, shorter and shorter pulses can be generated, and thereby a wider spectral bandwidth can be obtained \cite{Rotthardt2017,Kuhn2017,Toth2020,Kurucz2020,Rivas2017,Nagy2021,Krausz2009}. In the past few decades, the high-harmonic generation (HHG) process has become a widely used method to achieve coherent radiation in the extreme ultraviolet (XUV) region with the potential to produce electromagnetic pulses with subfemtosecond duration. HHG can be carried out on the surface of solids via the production of plasma mirrors  \cite{Mondal2018}, in liquids \cite{Luu2018}, in solid bandgap materials \cite{Nayak2019}, or in gaseous media \cite{Nayak2019,Corkum1993}. The most commonly used medium for HHG is a gaseous one, and in this case the generation process can be described by three steps \cite{Corkum1993}, which occur every half cycle of the driving electric field, resulting in distinct peaks in the generated XUV spectrum at odd multiples of the generating laser's central frequency. In this way the central wavelength of the high-order harmonics is fixed, and their spectral positions are defined by the generating laser source. The simple spectral tunability of these harmonic peaks can significantly widen the range of applications of these XUV sources \cite{Li2020}, e.g. in  chemical composition mapping \cite{Shapiro2014,Shi2019}, transient absorption spectroscopy \cite{Chang2020,Ding2021,Zurch2017,Yuki2019}, or XUV coherence tomography \cite{Wiesner2021,Nathanaael2019,Baksh2020}. 
\par
A known feature of high-harmonic generation is the inheritance of certain properties of the generating field \cite{Salaries1995,Huillier1991}; therefore, modulating the temporal evolution of the driving pulses or inducing spectral changes in the laser spectrum itself allows for the modification  of particular properties of the produced radiation. Recently, numerous experimental and theoretical studies have focused on the investigation of the impact of driving field modulation on the generated harmonics  \cite{Ciappina2012,Yavuz2012,Chou2015,Lee2001,Odzak2005}. Some of them focus on methods feasible for the extension of the cutoff  \cite{Chou2015}, while others discuss the possibility of the coherent control of the generated harmonics by chirped laser pulses  \cite{Lee2001}, and a few techniques applicable for the production of a HHG spectrum with a multiplateau structure \cite{Odzak2005,Xiang2009}. 
\par
Apart from the techniques summarized above, unconventional electric fields can be created as a sum of two temporally delayed ultrashort laser pulses \cite{Gulyas2020,Holzner,Gulyas2020,Raith2012,Perez2009}, whose application for HHG has a significant effect on the produced harmonics. Harmonic generation driven by double-pulse structures in a noncollinear focusing arrangement has been experimentally realized on numerous occasions. The HHG process in such a configuration has already been investigated, but the delay between the constituent pulses was much greater than the temporal duration of the pulses, and the aim was to optimize the HHG flux by preionizing the medium \cite{Daboussi2013}. Moreover, attempts have also been made to use three beams to generate harmonics in a noncollinear experimental layout, i.e. in BoxCARS geometry \cite{Negro2014}. The variation of the spectral position and the width of the generated harmonics was demonstrated, but it was attributed to the effects related to the spatiotemporal evolution of the laser field interacting with the medium. Here, in parallel to unveiling the real physical origin of the harmonic spectral changes observed in Ref. \cite{Negro2014}, we demonstrate a method that can be applicable in any attosecond pulse generation beamline for the quick and easy modification of the central photon energy of an XUV source.
\par
Following the method proposed theoretically in our previous work \cite{Gulyas2020}, a simple means to modulate the spectral properties of the generating laser beam is to exploit the spectral interference between two, temporally delayed, ultrashort laser pulses. When HHG is driven by these pulses, the photon energy of the generated harmonics can be tuned and adapted to a wide range of applications. This effect has recently been verified experimentally by Schuster \textit{et al.} \cite{schuster2021}, in parallel to our research. In their work however, the double pulses were produced by a Mach-Zehnder-type interferometer, in which case significant losses occur, since half of the invested laser energy is lost due to amplitude splitting. In the current work, we present an experimental validation of controlling harmonic photon energy by double generating pulses in a noncollinear generation geometry.
\par
The experiments are performed using the HR GHHG GAS beamline of the Extreme Light Infrastructure - Attosecond Light Pulse Source (ELI-ALPS) Research Facility \cite{Kuhn2017,Peng2020}. The high-harmonic generation beamline \cite{Peng2020} is seeded by one of the user-ready laser sources of ELI-ALPS \cite{Hadrich2016}: the high-repetition-rate (HR-1) laser system is specified to emit $1~\mathrm{mJ}$, sub-2-cycle pulses at $1030~\mathrm{nm}$ \cite{Hadrich2016}. The heart of the system is a powerful subpicosecond ytterbium fiber-based chirped  pulse amplifier (CPA). After the CPA, the pulses are postcompressed in a gas-filled hollow-core fiber \cite{Hadrich2016}. By reconfiguring the postcompression steps, one can reach a broad range of pulse durations ($6.2$ to $200~\mathrm{fs}$) , but the achievable average power drops as temporal duration decreases. Besides, fine-tuning of the laser power is also feasible in the long-pulse mode ($>18~\mathrm{fs}$). For the experiments reported in this Letter, we typically use a $30~\mathrm{fs}$ pulse duration and a power level of $90~\mathrm{W}$, which provide optimum, stable HHG in the current beamline geometry. Moreover, in this multicycle regime the carrier envelope phase (CEP) effects do not influence the HHG process, and the delay dependence of the individual harmonic peaks can be easily tracked.

\begin{figure}[t]
    \includegraphics[width=1\linewidth]{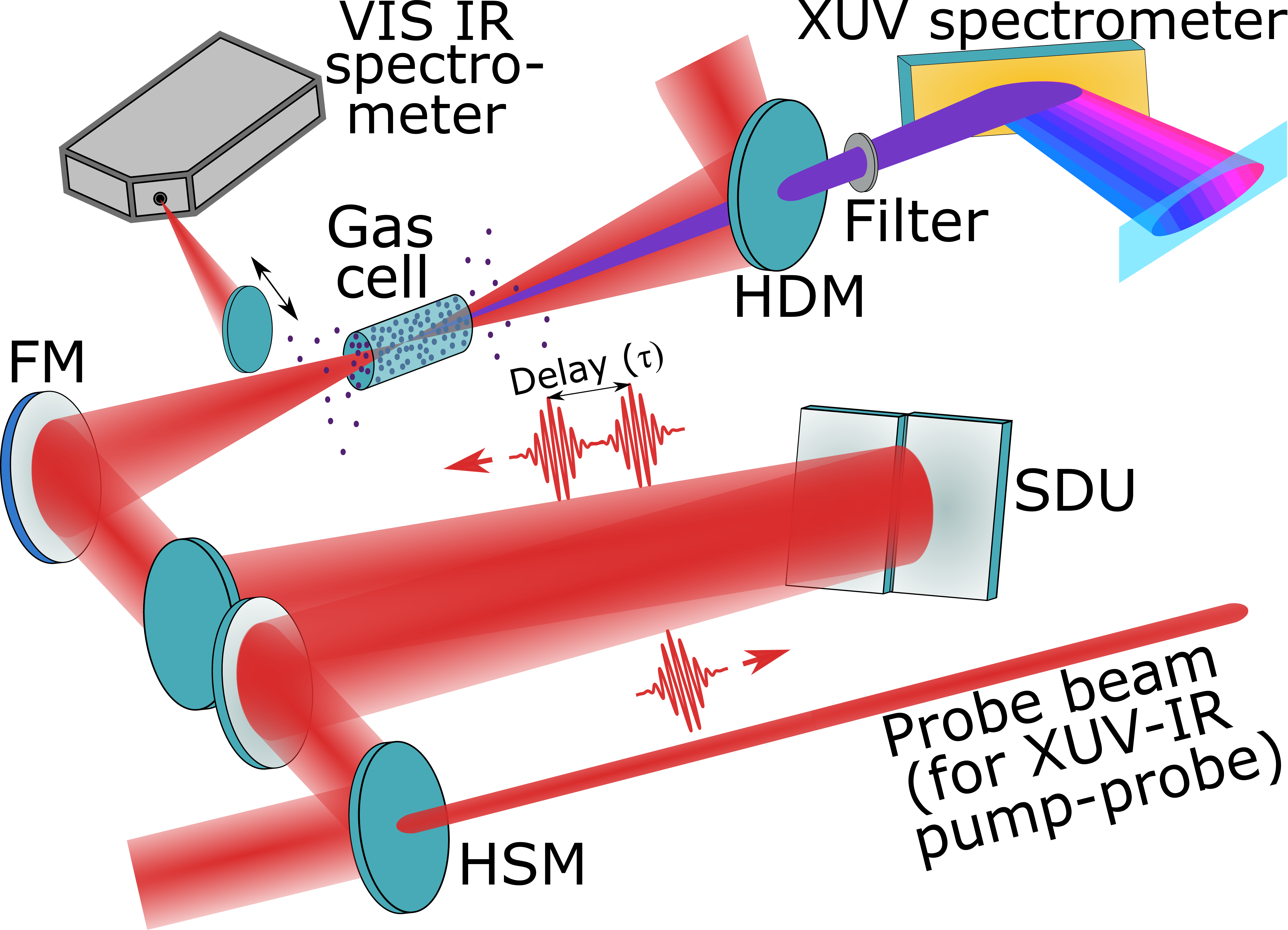}
    \caption{Schematic experimental setup of the HR GHHG GAS attosecond beamline at ELI-ALPS. The main optical components of the setup are: holey-splitting mirror (HSM), split-and-delay unit (SDU), focusing mirror (FM), visible-infrared (VIS IR) spectrometer, gas cell, holey-dump mirror (HDM), aluminum filter, and XUV flat-field spectrometer. Red represents the IR beam, and purple the generated XUV beam.}
    \label{fig:expset}
\end{figure}

\par
Compared with the beamline configuration detailed by Ye \textit{et al.} in ref. \cite{Peng2020} only a minor modification was made on the beamline (Fig.  \ref{fig:expset}): one of the steering mirrors is replaced by a custom-made SDU \cite{Campi2016}, while the other optics remain unchanged. A simplified schematic layout of the beamline is given in Fig. \ref{fig:expset}. The incoming laser beam is split by the holey-splitting mirror (HSM) to an annular part and a central part, which are used for HHG and as a probe beam for XUV IR pump-probe experiments, respectively. After the split, the annular part is divided into two halves by wave-front splitting and, at the same time, the SDU introduces a pulse-duration-comparable temporal delay ($\tau$) between them. The two half beams propagate together and are focused by the same focusing mirror (FM) into the generation medium where, with the proper alignment of the SDU (both spatial and temporal overlaps can be finely adjusted), the focused beams overlap spatially and a double-pulse structure appears. The modulated laser spectra are recorded before the generation cell by inserting a flip-mirror steering the beams towards a VIS-IR spectrometer. The generated XUV radiation propagates through the hole of another holey mirror, the holey-dump mirror (HDM), while the generating IR is reflected and dumped. The residual IR is eliminated by an aluminum filter. After further propagation through XUV optics \cite{Peng2020}, the HHG beam reaches the XUV spectrometer section, consisting of an XUV grating, a microchannel plate, a phosphor screen and a CMOS camera.

\begin{figure}[b]
    \includegraphics[width=1\linewidth]{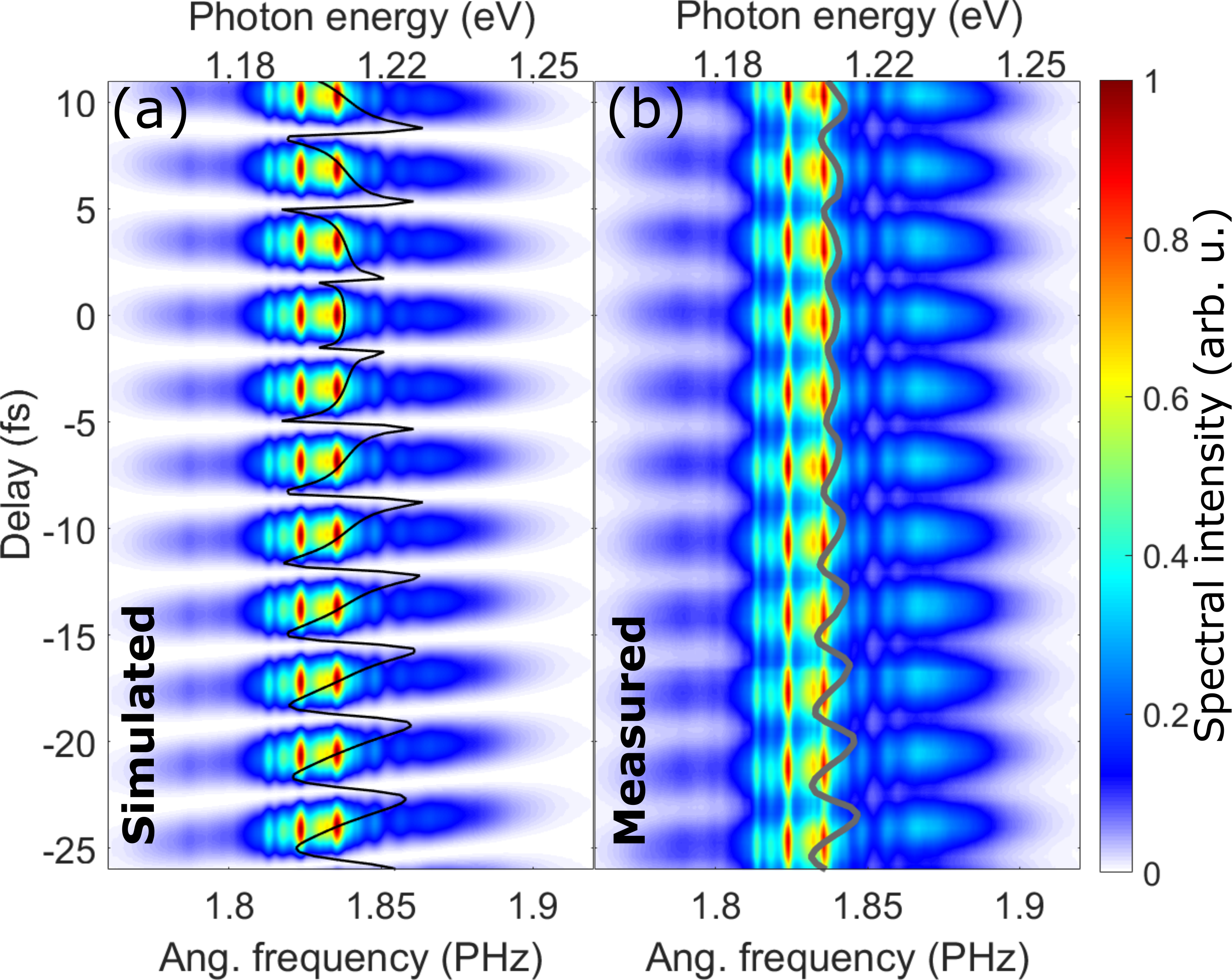}
    \caption{Simulated (a) and experimentally measured (b) spectra of the temporal double-pulse structure as a function of the time delay between the constituting pulses. The black (a) and gray (b) curves show the delay dependence of the central angular frequency.}
    \label{fig:spectrum}
\end{figure}

To demonstrate the tunability of the laser's central wavelength, simulated spectral interferograms are presented in Fig. \ref{fig:spectrum}.a as a function of time separation between the two constituting pulses (Fig. \ref{fig:spectrum}.a, colored map), having the same temporal amplitude and assuming a pointlike laser source. To simulate double pulses the measured laser spectrum is used. We can clearly see the strong modulation of the laser spectrum, whereby the central angular frequency periodically changes (Fig. \ref{fig:spectrum}.a, black curve), calculated by the formula $\omega_0(\tau) = \int_{-\infty}^{\infty} \omega I(\omega,\tau) d\omega/\int_{-\infty}^{\infty} I(\omega,\tau) d\omega$ at each delay, where $I(\omega,\tau)$ is the recorded spectral intensity. The same periodicity is visible in the spectrum of the experimentally produced double pulses (Fig. \ref{fig:spectrum}.b, colored map), where the corresponding central angular frequency is shown by the gray curve. The period of the black and

\onecolumngrid

\begin{center}
\begin{figure*}[t]
    \includegraphics[width=1\linewidth]{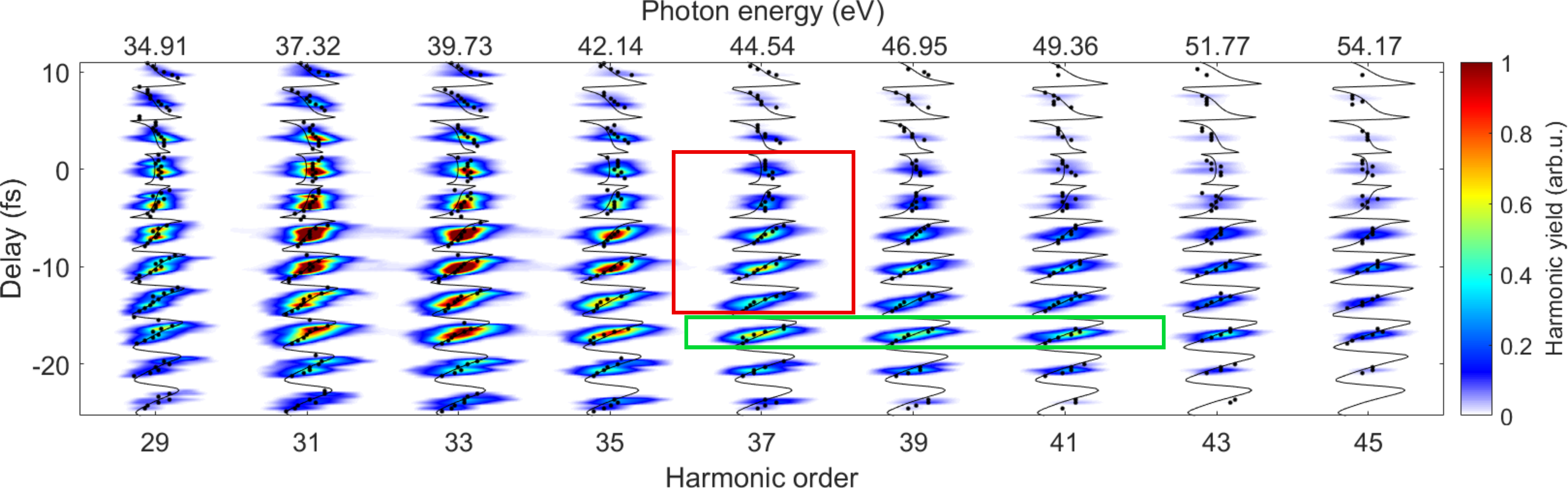}
    \caption{Experimentally generated high-order harmonic spectra as a function of the delay between the driving pulses (colored map). The black dots represent the exact spectral positions of the harmonic peaks. The black curves show where each harmonic should appear according to the central angular frequency of the generation field, i.e. according to the black curve of Fig. \ref{fig:spectrum}.a.\vspace{-8mm}}
    \label{fig:hhgfull}
\end{figure*}
\end{center}

\twocolumngrid

\noindent
gray curves is the optical cycle of the applied laser, which is approximately $3.4~\mathrm{fs}$ in our case. The oscillation of the central angular frequency extracted from the simulations (black curve) is more explicit than that of the curve directly calculated from the experimentally recorded data (gray curve), which can be attributed to the generation geometry and detection configuration. After focusing, the two half beams propagate at an angle of approximately $20~\mathrm{mrad}$ and they overlap spatially only in a small region around the focus, where the laser interacts with the target gas (the periodicity of the spatial interference is of the order of the focused beam size). For this reason, the two half beams enter the spectrometer neither in equal proportions nor at the same angle of incidence, and thus the visibility of the spectral interference decreases. 
\par
The periodic structure of the central angular frequency visible in the driving laser (Fig. \ref{fig:spectrum}) appears also in the generated high-order harmonics (Fig. \ref{fig:hhgfull}). Fig. \ref{fig:hhgfull}. shows a delay scan, where all experimental conditions are unchanged, except for the temporal delay between the constituting pulses of the double-pulse structure. A typical scan takes 15 minutes. However, for long-term measurements, it is required to actively stabilize the overlap and delay of the two half beams in the focus \cite{Campi2016}. Harmonics are generated in argon gas at $210~\mathrm{mbar}$ static pressure in a 4-mm-long generation gas cell. The laser pulses are focused by a mirror with a focal length of  $900~\mathrm{mm}$. Due to the loose focusing geometry (the Rayleigh range is ~10 mm), with the current laser parameters the achievable peak intensity can only exceed the ionization limit when the two half beams are in proper spatial and temporal overlap. Since changes in the time separation between the generating pulses cause either destructive or constructive interference between the pulses, the laser peak intensity periodically falls below the ionization threshold. As a result, XUV radiation periodically appears and disappears as a function of delay. The phase-matching conditions are optimized at a delay where we assume a perfect temporal overlap. The absolute zero delay of the scan is determined based on the central angular frequency curve (black curve), which has a well-defined zero delay position. We attribute asymmetry at zero time delay to the imperfections of the double-pulse structure, which influence the phase-matching conditions. The black curves at each harmonic order in  Fig. \ref{fig:hhgfull} are the expected positions of the harmonics, determined by multiplying the central angular frequency of the laser spectra ($E_q(\tau) = q \cdot \hbar \cdot \omega_0(\tau)$, where $E_q(\tau)$ is the harmonic photon energy, $\hbar$ is the reduced Planck constant, q is the harmonic order, and $\omega_0(\tau)$ is the central angular frequency of the laser). The black dots in the figure represent the exact positions of the harmonic peaks. It can be clearly observed that the positions of the harmonic peaks follow what is expected from the change of the central angular frequency of the generating laser field.
\par
\begin{figure}[t]
    \includegraphics[width=1\linewidth]{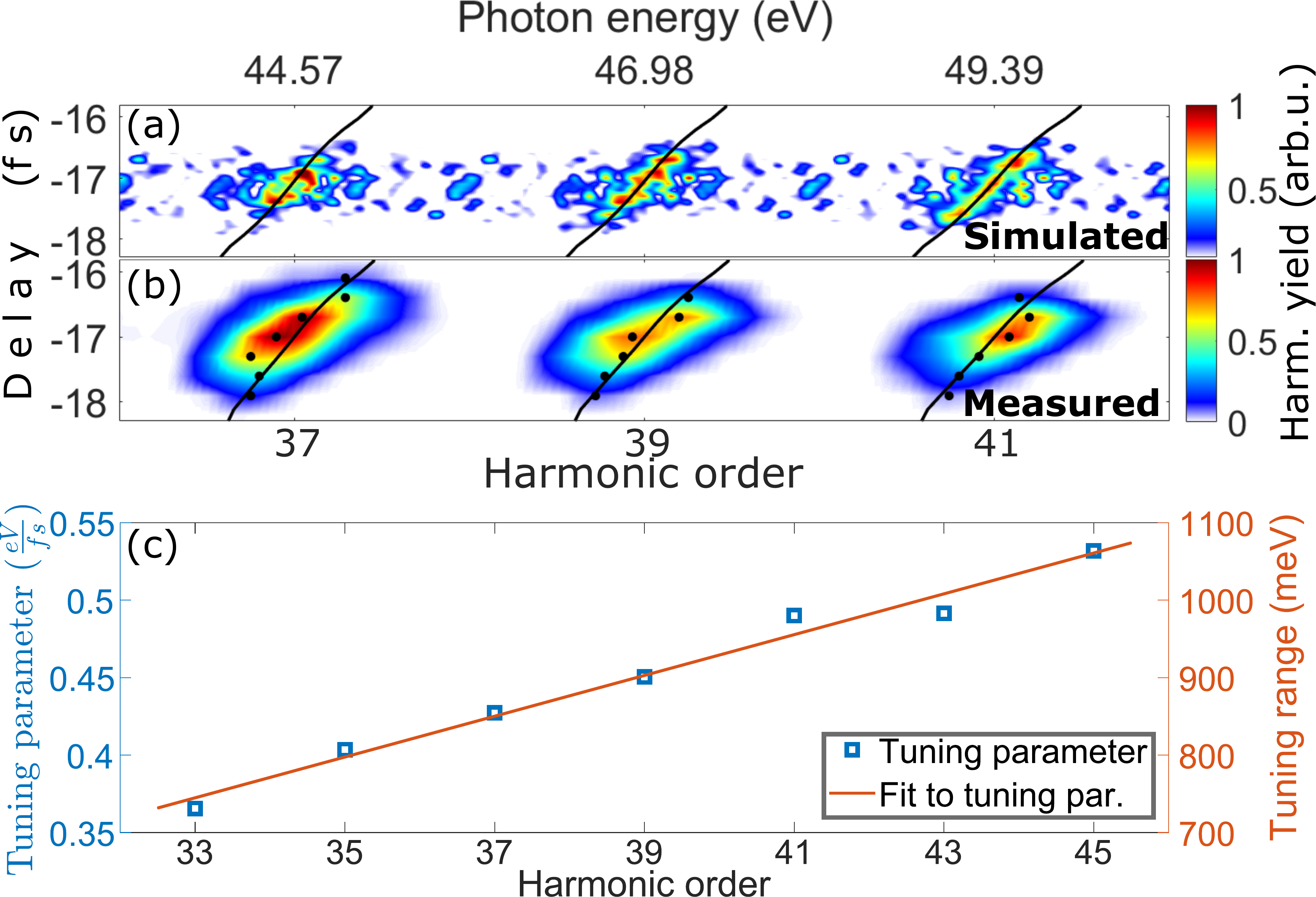}
    \caption{(a) Simulated and (b) experimentally measured harmonics from the 37th to 41st harmonic orders around $-17~\mathrm{fs}$ delay. The black curves demonstrate where the harmonic peaks should appear according to the central angular frequency of the driving laser. The black dots in (b) show the measured positions of the harmonic peaks. (c) Demonstration of the achievable tuning ranges of different harmonics.\vspace{-2mm}}
    \label{fig:energytune}
\end{figure}
\begin{figure}[t]
    \includegraphics[width=1\linewidth]{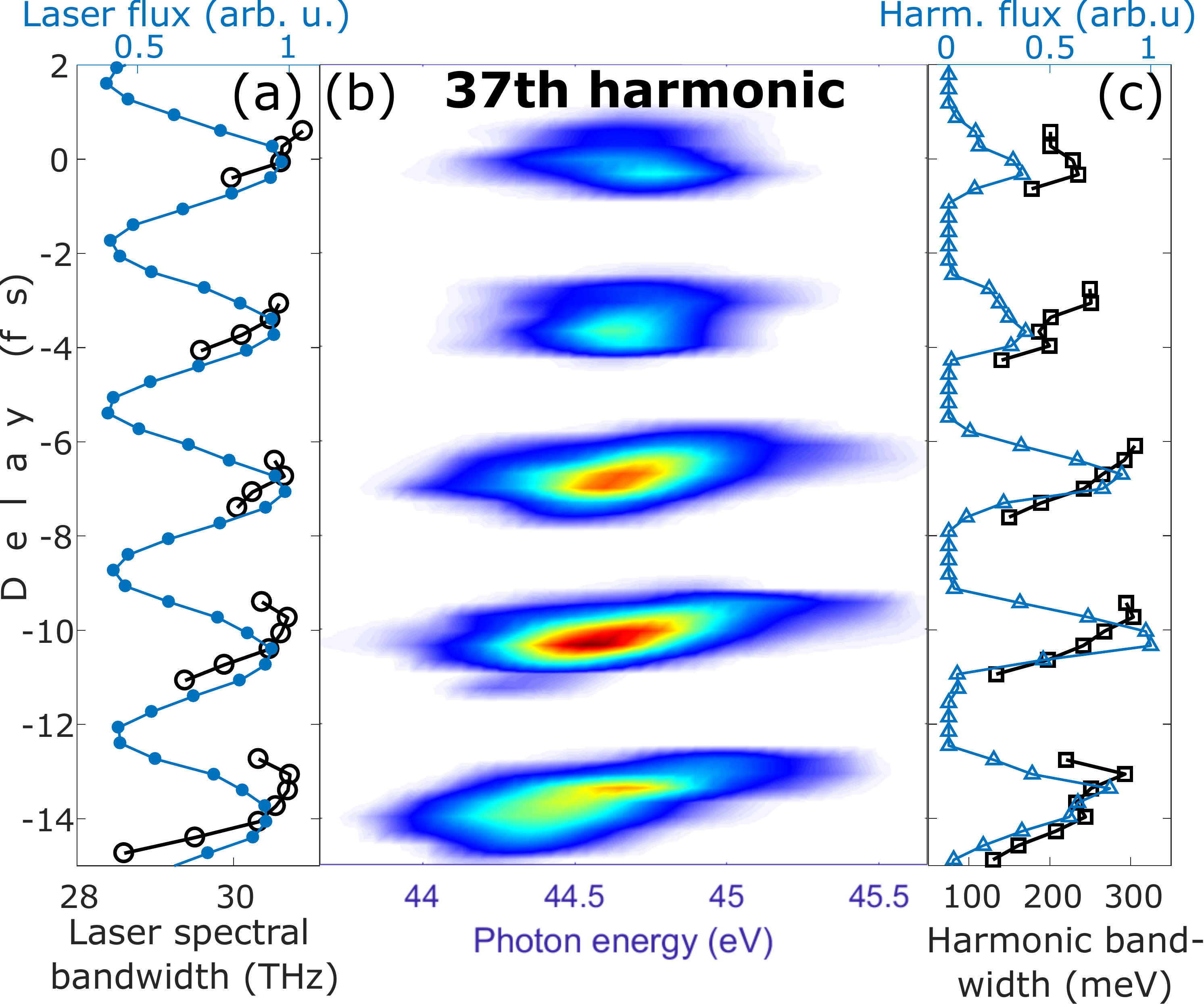}
    \caption{(a) Spectral bandwidth variation (black curve with open circles) and the flux (blue curve with filled circles) of the measured driving laser spectra as a function of the delay. (b) Th 37th harmonic on a wide delay range showing the positions, where the driving pulses interfere constructively. (c) The spectral bandwidth variation (black curve with squares) and flux (blue curve triangles) of the 37th harmonic.\vspace{-2mm}}
    \label{fig:spnarrow}
\end{figure}

For a more detailed investigation, we choose specific harmonics (37th to 41st) in a delay range (from {$-18.3$ to $-15.8~\mathrm{fs}$}) where the two driving pulses interfere constructively, as shown in Fig. \ref{fig:energytune}. For comparison, Fig. \ref{fig:energytune}.a shows single-atom simulation results, using the driving laser spectrum in Fig. \ref{fig:spectrum}.a, and applying $100~\mathrm{as}$ resolution in the delay. In the experiment, the interaction length is confined to a small region where the two pulses overlap spatially and temporally, so the single-atom model of the laser-matter interaction is used to calculate the dipole spectra \cite{Lewenstein1994}, without considering the propagation processes such as plasma generation, self-focusing, absorption and dispersion effects. In the simulation the target gas is argon, which has an ionization potential of $15.76~\mathrm{eV}$. The intensity of each pulse is $1.2 \times 10^{14}~\mathrm{W/cm^2}$ and the spectral phase, including CEP, is the same. Fig. \ref{fig:energytune}.b is the enlarged green quadrant indicated in Fig. \ref{fig:hhgfull}. The black curves in Fig. \ref{fig:energytune}.a and b illustrate where the harmonics should appear according to the driving central angular frequency. In Fig. \ref{fig:energytune}.b the black dots show the exact positions of the experimentally measured harmonic peaks. By calculating the $dE_q(\tau)/d\tau$ tuning parameter (Fig. \ref{fig:energytune}.c, blue squares) the exact tuning ranges of different harmonic orders can be determined based on a linear fit (Fig. \ref{fig:energytune}.c, orange line), showing the expected linear increase of the tuning range with increasing harmonic order \cite{Gulyas2020,schuster2021}. Assuming a $\tau_q = 2~\mathrm{fs}$ delay range, where the harmonics are visible (see Fig. \ref{fig:spnarrow}.c with flux as a function of delay), the exact tuning ranges ($dE_q(\tau)/d\tau \times \tau_q)$ are shown by the right-hand axis of Fig. \ref{fig:energytune}.c. The presented tuning range can be further increased by performing HHG with higher delays, in which case the slope of the fitted curve (similar to Fig. \ref{fig:energytune}.c, orange line) would be larger. It is important to note that the tuning range of the highest-order harmonic, presented in Fig. \ref{fig:energytune}.c, is close to the photon energy of the driving laser, which is $1200~\mathrm{meV}$ in our case (see Fig. \ref{fig:spectrum}).

\par
Fig. \ref{fig:spnarrow}.a demonstrates the spectral bandwidth variation of the driving laser source (black curve with open circles). This is calculated by the formula $\Delta\omega(\tau)^2 = \int_{-\infty}^{\infty} (\omega-\omega_0(\tau))^2  I(\omega,\tau)  d\omega/\int_{-\infty}^{\infty} I(\omega,\tau) d\omega$ using the spectral interferograms in Fig. \ref{fig:spectrum}.b, with $\omega_0(\tau)$ being the central angular frequency (Fig. \ref{fig:spectrum}.b, gray curve) and $I(\omega,\tau)$ being the laser spectral intensity. The blue curve with filled circles in Fig. \ref{fig:spnarrow}.a represents the laser flux. Fig. \ref{fig:spnarrow}.b shows the investigated harmonics taken from the red quadrant of Fig. \ref{fig:hhgfull}. Fig. \ref{fig:spnarrow}.c shows the bandwidth variation (black curve with squares) and the flux (blue curve with triangles) of the 37th harmonic as a function of delay. As can be seen in Fig. \ref{fig:spnarrow}.c, the harmonic flux shows narrower peaks as a function of delay, in accordance with the approximately fourth-power dependence of harmonic intensity on the laser intensity \cite{Lewenstein1994}. In terms of the bandwidth curve, $\omega_0(\tau)$ means the harmonic peak position and $I(\omega,\tau)$ is the harmonic spectral intensity in the used formula. The bandwidth of the harmonics as a function of time delay has similar tendencies to the bandwidth of the driving laser. A moderate narrowing effect can be observed in Fig. \ref{fig:spnarrow}.a for the driving laser field, especially at higher delays. However, this does not appear in Fig. \ref{fig:spnarrow}.c; instead the harmonic bandwidth varies around the bandwidth values measured at zero delay. We attribute this to the driving laser’s nonideal spectral amplitude, which includes intense sharp peaks (Fig. \ref{fig:spectrum}), leading to differences compared with the earlier theoretical results, and thus predicting a clear harmonic bandwidth narrowing effect with increasing delay \cite{Gulyas2020}. When shorter driving pulses with smoother spectral amplitudes are used, the generated harmonics are expected to inherit the bandwidth characteristics of the driving laser more clearly \cite{Gulyas2020,Negro2014}. Nevertheless, by combining this setup with an XUV monochromator \cite{Luca2009,Fabio2017}, which allows for fine bandwidth selection typically down to $100~\mathrm{meV}$ or below \cite{Luca2014}, we can realize an XUV source with easily and separately adjustable central photon energy and bandwidth.

\par
In the current work we experimentally investigate the HHG process in argon gas, driven by double pulses having a pulse duration comparable to the delay. By performing these measurements, we prove the theoretically predicted result \cite{Gulyas2020}, i.e. by changing the delay between the pulses constituting the double-pulse structure, the photon energy of the generated harmonics can be controlled. A great advantage of the proposed method is easy implementation in any attosecond pulse generation beamline, regardless of the driving laser source. Moreover, by using laser sources emitting shorter pulses with a broader spectral bandwidth, photon energy can be tuned in a wider energy range \cite{Gulyas2020}. In cases where few-cycle driving laser pulses are used and higher harmonic orders are investigated, the entire spectral region up to the cutoff can be covered. Furthermore, by applying a split-and-delay unit to produce the double-pulse structure using wave-front splitting instead of amplitude splitting \cite{schuster2021}, almost the total invested energy can be utilized for harmonic generation. By combining this arrangement with an XUV monochromator \cite{Luca2009,Fabio2017}, we can build an XUV source that allows for the quick and simple modification of the central photon energy and bandwidth to satisfy different experimental needs and open the way towards a wider range of applications.
\par
This research is financially supported by the ELI-ALPS project (No. GINOP-2.3.6-15-2015-00001), which is funded by the European Union and co-financed by the European Regional Development Fund. 

\bibliographystyle{ieeetr}
\bibliography{References}

\end{document}